\documentstyle[11pt]{article}

\begin{document}

\newcommand{\beq}{\begin{equation}}
\newcommand{\eeq}{\end{equation}}
\newcommand{\bea}{\begin{eqnarray}}
\newcommand{\eea}{\end{eqnarray}}

\newcommand{\chii}{\raise.5ex\hbox{$\chi$}}
\newcommand{\Z}{Z \! \! \! Z}
\newcommand{\R}{I \! \! R}
\newcommand{\N}{I \! \! N}
\newcommand{\C}{I \! \! \! \! C}

\newcommand{\noi}{\noindent}
\newcommand{\vs}{\vspace{5mm}}
\newcommand{\ie}{{${ i.e.\ }$}}
\newcommand{\eg}{{${ e.g.\ }$}}
\newcommand{\ea}{{${ et~al.\ }$}}
\newcommand{\hf}{{\scriptstyle{1 \over 2}}}
\newcommand{\ih}{{\scriptstyle{i \over \hbar}}}
\newcommand{\hi}{{\scriptstyle{ \hbar \over i}}}
\newcommand{\itwoh}{{\scriptstyle{i \over {2\hbar}}}}
\newcommand{\dbrst}{\delta_{BRST}}

\newcommand{\deder}[1]{{ 
 {\stackrel{\raise.1ex\hbox{$\leftarrow$}}{\delta^r}   } 
\over {   \delta {#1}}  }}
\newcommand{\dedel}[1]{{ 
 {\stackrel{\lower.3ex \hbox{$\rightarrow$}}{\delta^l}   }
 \over {   \delta {#1}}  }}
\newcommand{\dedetwo}[2]{{    { \delta {#1}} \over {   \delta {#2}}  }}
\newcommand{\dedetre}[3]{{ \left({ \delta {#1}} \over {   \delta {#2}}  
 \right)_{\! \! ({#3})} }}

\newcommand{\papar}[1]{{ 
 {\stackrel{\raise.1ex\hbox{$\leftarrow$}}{\partial^r}   } 
\over {   \partial {#1}}  }}
\newcommand{\papal}[1]{{ 
 {\stackrel{\lower.3ex \hbox{$\rightarrow$}}{\partial^l}   }
 \over {   \partial {#1}}  }}
\newcommand{\papatwo}[2]{{   { \partial {#1}} \over {   \partial {#2}}  }}
\newcommand{\papa}[1]{{  {\partial} \over {\partial {#1}}  }}
\newcommand{\papara}[1]{{ 
 {\stackrel{\raise.1ex\hbox{$\leftarrow$}}{\partial}   } 
\over {   \partial {#1}}  }}

\newcommand{\ddr}[1]{{ 
 {\stackrel{\raise.1ex\hbox{$\leftarrow$}}{\delta^r}   } 
\over {   \delta {#1}}  }}
\newcommand{\ddl}[1]{{ 
 {\stackrel{\lower.3ex \hbox{$\rightarrow$}}{\delta^l}   }
 \over {   \delta {#1}}  }}
\newcommand{\dd}[1]{{  {\delta} \over {\delta {#1}}  }}
\newcommand{\pa}{\partial}
\newcommand{\sokkel}[1]{\!  {\lower 1.5ex \hbox{${\scriptstyle {#1}}$}}}  
\newcommand{\larrow}[1]{\stackrel{\rightarrow}{#1}}
\newcommand{\rarrow}[1]{\stackrel{\leftarrow}{#1}}
\newcommand{\twobyone}[2]{\left(\begin{array}{c}{#1} \cr
                                   {#2} \end{array} \right)}
\newcommand{\twobytwo}[4]{\left[\begin{array}{ccc}{#1}&&{#2} \cr
                                  {#3} && {#4} \end{array} \right]}
\newcommand{\fourbyone}[4]{\left(\begin{array}{c}{#1} \cr{#2} \cr{#3} \cr
                                   {#4} \end{array} \right)}

\newcommand{\eq}[1]{{(\ref{#1})}}
\newcommand{\Eq}[1]{{eq.~(\ref{#1})}}
\newcommand{\Ref}[1]{{Ref.~\cite{#1}}}
\newcommand{\mb}[1]{{\mbox{${#1}$}}}
\newcommand{\equi}[1]{\stackrel{{#1}}{=}}

\newcommand{\proofbox}{\begin{flushright}
${\,\lower0.9pt\vbox{\hrule \hbox{\vrule
height 0.2 cm \hskip 0.2 cm \vrule height 0.2 cm}\hrule}\,}$
\end{flushright}}


\begin{titlepage}

\title{
\normalsize
\rightline{UFIFT-HEP-00-18}
\rightline{hep-th/0007192}
\vspace{2.5 cm}
\Large\bf A Note on Non-Locality and Ostrogradski's Construction\\ 
}

\author{{\sc K.~Bering}
\footnote{Email address: {\tt bering@phys.ufl.edu, bering@nbi.dk}}
\\
Institute for Fundamental Theory
\footnote{Address after Oct.\ 1st, 2000: Rockefeller University,
Department of Physics, Box 254, 1230 York Avenue, 
New York, NY, 10021-6399, USA.}
\\ Department of Physics\\
University of Florida\\Florida 32611, USA\\
}

\date{July 2000}

\maketitle
\begin{abstract}
We consider the Hamiltonian treatment of non-local theories and 
Ostrogradski's formalism. This has recently also been discussed by 
Woodard (hep-th/0006207) and by Gomis, Kamimura and Llosa (hep-th/0006235).  
In our approach we recast the second class constraints into first
class constraints and invoke the boundary Poisson bracket.
\end{abstract}

\bigskip

\vspace*{\fill}

\noi
PACS number(s): 02.70.Pt, 03.50.-z, 04.20.Fy, 11.10.Ef, 11.10.Lm, 45.20.Jj.

\end{titlepage}
\vfill
\newpage

\setcounter{equation}{0}
\section{Motivations}

\vs
\noi
Woodard's construction \cite{woodardnew} has boundaries 
in the non-local integration range at \mb{x=0} and \mb{x=\Delta t}. 
If caution is not applied, a breakdown of the Jacobi identity
and a loss of differentiability of functionals may lead
to inconsistencies. As mentioned in \cite{woodardnew}, additional
boundary conditions must be imposed. Intuitively, the boundaries of
the non-local integration range should not have any physical 
significance. We shall devise a slightly modified
construction that avoids boundaries and hence steers clear of such hazards. 

\vs
\noi
In addition, we are interested in the close relationship between
Ostrogradski's derivatives and the higher Euler-Lagrange derivatives
used among other things to construct boundary Poisson brackets, 
cf.\ \cite{soloviev,bering,solocomp,family}.

\vs
\noi
Compared with earlier work, we do not have second class constraints. 
We obtain a clear separation of the equations of motion from the
constraints, which is preferable at the conceptual level.

\vs
\noi 
We shall have nothing to say about the soundness of non-local
theories and higher derivative theories in general. Instead we refer to the
ongoing discussion in the Literature \cite{woodardnew,woodardold}.

\setcounter{equation}{0}
\section{Lagrangian Variables}
\label{seclagrvar}

\vs
\noi
{}For simplicity, we shall consider a \mb{0+1} dimensional systems with one
dynamical variable \mb{q(t)}. The construction can be straightforwardly 
extended to include more variables and to field theories. We shall
assume that the time coordinate \mb{t} has no temporal boundaries, 
\ie \mb{t \in \R} can take any real value. 

\vs
\noi
We are free to think \cite{llosaold} of the dynamical variable \mb{q(t)} as 
a \mb{1+1} dimensional field \mb{Q(x,t)} that satisfies a chirality condition
\beq
{{d} \over {dt}}Q(x,t)~=~ \pa_{x} Q(x,t)~.
\label{qxt}
\eeq
Explicitly, the one-to-one correspondence between \mb{q(t)} and 
the left-mover \mb{Q(x,t)} is given by \mb{Q(x,t)=q(x+t)}. 
Keep in mind for later that \Eq{qxt} implies
\beq
\left({{d} \over {dt}}\right)^{n} Q(x,t)~=~ \left(\pa_{x}\right)^{n}
Q(x,t)~,~~n \in \N_{0}~.
\label{mqxt}
\eeq

\setcounter{equation}{0}
\section{Lagrangian}

\vs
\noi
By {\em non-locality} we mean that the Lagrangian \mb{L[q](t)} depends on 
the dynamical variable \mb{q} at other times than \mb{t}.
To deal with this in a systematic manner we shall assume that the
Lagrangian can be written as a $d$-dimensional integral
\beq
 L[Q](t)~=~\int_{-\infty}^{\infty} \! \! \! \! d x^{1}\cdots 
\int_{-\infty}^{\infty} \! \! \! \! d x^{d} ~{\cal L}(x^{1},\ldots,x^{d},t)
\label{lint}
\eeq
over a density function \mb{{\cal L}}. To be precise, besides the 
explicit dependence of \mb{x^{1},\ldots,x^{d}} and \mb{t}, 
the density function \mb{{\cal L}(x^{1},\ldots,x^{d},t)}
is assumed to be a function of a {\em finite} number 
of the following variables:
\beq
(\pa_{x})^{k} Q(x^{i},t)~,~~i=1,\ldots,d,~,~~k \in \N_{0}~.
\label{qlist}
\eeq

\vs
\noi
The replacement of \mb{q(t)} with \mb{Q(x,t)} has several advantages:

\vs
\noi
{}First of all, \mb{L[Q](t)} does only depend on \mb{Q(x,t)}'s of
the very same \mb{t}. 
The non-locality is encoded in the new variable \mb{x}.
Negative and positive values of \mb{x} correspond to interactions 
with the past and the future, respectively. (We should stress 
that \mb{x} has nothing to do with space; we have merely named the new 
variable \mb{x} because the formulas fit the framework of field theory.)

\vs
\noi
Secondly, in the \mb{Q}-formulation we have removed all derivatives wrt.\
\mb{t} which appeared in the original Lagrangian \mb{L[q](t)} by using the 
chirality condition \eq{mqxt}. This will prepare us for a very 
smooth transition into its Hamiltonian counterpart, \ie the Lagrangian
does not depend on the velocities,  on the accelerations, etc.\ and 
hence on the momenta.
(Note that in the Hamiltonian formulation our starting point will be 
\mb{L[Q](t)} {\em without} assuming chirality of \mb{Q(x,t)}. We shall 
later see that in the Hamiltonian formulation the chirality condition 
\eq{qxt} becomes the equation of motion for \mb{Q(x,t)}.)

\setcounter{equation}{0}
\section{Local Field Theory}
\label{lft}

\vs
\noi
A functional \eq{lint} with the assumption that its density
\mb{{\cal L}(x^{1},\ldots,x^{d},t)} depends on a
{\em finite} number of variables from the list \eq{qlist} is
commonly known as the very definition of a {\em local} functional.
We may say that the original {\em non-local} theory has become 
a {\em local field} theory.

\vs
\noi
The case of {\em discrete} non-locality has been studied 
extensively in the Literature \cite{woodardnew,woodardold}. 
We define it as the case where there exists a discrete family 
of curves \mb{\{t \mapsto x_{i}(t)\}_{i \in I}} 
so that \mb{L[Q](t)} only depends on a {\em finite} number
of the following variable: \mb{t}, \mb{x_{i}(t)} and
\beq
(\pa_{x})^{k} Q(x_{i}(t),t)~,~~i \in I~~,~~k \in \N_{0}~.
\label{dqlist}
\eeq
{}For technical reasons we shall assume the function
\mb{{\cal L}(x^{1},\ldots,x^{d},t)} is \mb{C^{\infty}}-differentiable. 
This smoothness assumption unfortunately rules out the case of 
discrete non-locality. However, we shall investigate the
discrete case in Section~\ref{dica}.

\setcounter{equation}{0}
\section{Compact Support}
\label{compact}

\vs
\noi
We assume that the density function \mb{{\cal L}} (and other physically 
meaningful objects) has a {\em compact} support in the
\mb{x}-directions. As we shall see this assumption has radical
implications for the theory.

\setcounter{equation}{0}
\section{Momenta}
\label{secp}

\vs
\noi
In a non-local Lagrangian theory the usual definition of momenta as the
derivatives of the Lagrangian wrt.\ the velocities is not useful. 
Instead, we shall seek a new and better definition.
We take as our initial guess the partial differential equation
\beq
{{d} \over {dt}}P(x,t)~=~ \pa_{x} P(x,t)
+{{\delta L [Q] (t)} \over {\delta Q^{(0)}(x,t)}}~,
\label{pdiff}
\eeq 
subjected to the following boundary condition:
\beq
 P(x,t)~ {\rm has~compact~support~in~the}~x{\rm -direction}.
\label{lpbc}
\eeq
The \mb{(0)} appearing on the symbol for the functional derivatives 
denotes that we use the algebraic Euler-Lagrange definition rather
than a variational definition of the functional derivatives, 
cf.\ \cite{bering}. We shall see below that our initial guess
\eq{pdiff} needs off-shell modifications.

\vs
\noi
Both the partial differential equation \eq{pdiff} and the boundary 
condition \eq{lpbc} appear naturally in the Hamiltonian treatment 
to be given below. 
Purely from a Lagrangian perspective, the boundary condition \eq{lpbc}
arises as a natural consequence of requiring the density function
\mb{{\cal L}} itself to have compact support.

\vs
\noi
In general the above boundary value problem \eq{pdiff} and \eq{lpbc} 
may not have continuous solutions.
Let us allow for a potential discontinuity along a curve 
\mb{x=x_{0}(t)}. The unique solution is then \cite{llosaold}
\bea
P(x,t)&=&\int_{-\infty}^{\infty} \! \! \! \! d x' 
\int_{-\infty}^{\infty}\! \! \! \!d t'~
\delta_{\R}(x'\!+\!t'\!-\!x\!-\!t)
\left[\theta(x'\!-\!x) ~ \theta(x\!-\!x_{0}(t))
-\theta(x_{0}(t)\!-\!x)~ \theta(x\!-\!x') \right] 
{{\delta L [Q] (t')} \over {\delta Q^{(0)}(x',t')}}\cr\cr 
&=&\int_{-\infty}^{\infty}\! \! \! \! d u ~
\left[ \theta(x\!-\!x_{0}(t))~\theta(u) 
-\theta(x_{0}(t)\!-\!x)~\theta(-u) \right] 
{{\delta L [Q] (t\!-\!u)} \over {\delta Q^{(0)}(x\!+\!u,t\!-\!u)}}~. 
\label{pdef}
\eea
In a Lagrangian treatment we promote this formula  
to be the very definition of momenta (see \cite[formula
(20)]{woodardnew} and \cite[formula (6)]{llosanew} ).
Needless to say, formula \eq{pdef} may have other
discontinuities if  \mb{\delta L [Q](t)/ \delta Q^{(0)}(x,t)} 
is singular, cf.\ Section~\ref{dica}.

\setcounter{equation}{0}
\section{Lagrangian Equation of Motion}

\vs
\noi
Let \mb{S[q] =\int_{-\infty}^{\infty} \! \! dt ~L[q](t)} denote the action. 
Recall that we cannot vary \mb{Q} freely because of the chirality 
condition \Eq{qxt}. Therefore, the Lagrangian Equation of Motion 
for \mb{Q} does not provide us with the relevant physical information.
Instead, the pertinent equation of motion is given by the Lagrangian 
Equation of Motion for \mb{q}:
\bea
0&=&{{\delta S[q]} \over {\delta q^{(0)}(x\!+\!t)}}
~\equi{\eq{qxt}}~\int_{-\infty}^{\infty}\! \! \! \! d u ~
{{\delta L [Q] (t\!-\!u)} \over {\delta Q^{(0)}(x\!+\!u,t\!-\!u)}} \cr\cr
&\equiv &\int_{-\infty}^{\infty} \! \! \! \! d x' 
\int_{-\infty}^{\infty}\! \! \! \!d t'~
\delta_{\R}(x'\!+\!t'\!-\!x\!-\!t)~
{{\delta L [Q] (t')} \over {\delta Q^{(0)}(x',t')}}~. 
\label{leom}
\eea
In a \mb{Q}-formulation, with the chirality condition 
\Eq{qxt} not imposed manifestly, the Equation of Motion for 
\mb{q} does strictly speaking  not make sense (at least not a priori). 
However, even in that case, it is natural to call the last two 
expressions in \Eq{leom} the Lagrangian Equation of Motion for \mb{q}.

\vs
\noi
Note that \Eq{leom} is precisely the condition that the momentum  formula 
\eq{pdef} does not have an extra  discontinuity at \mb{x=x_{0}(t)}:
\beq
\lim_{x \to x_{0}(t)^{-}} P(x,t)~=~\lim_{x \to x_{0}(t)^{+}} P(x,t)~.
\eeq
{}From a Hamiltonian point of view, where the momentum is a fundamental
rather than a derived quantity, this is of course trivially guaranteed
by restricting ourselves to continuous fields. 
(In fact we shall only allow  \mb{C^{\infty}}-fields.)

\vs
\noi
Closely related to this fact is the following: If we include singular 
terms at the kink curve, the above momentum formula \eq{pdef} 
satisfies the following partial differential equation:
\beq
{{d} \over {dt}}P(x,t)~=~ \pa_{x} P(x,t)
+{{\delta L [Q] (t)} \over {\delta Q^{(0)}(x,t)}}
+\left(\dot{x}_{0}(t)\!-\!1 \right)~\delta_{\R}(x\!-\!x_{0}(t))
~{{\delta S[q]} \over {\delta q^{(0)}(x\!+\!t)}} ~.
\eeq
This differs from the original \Eq{pdiff} with a delta function 
contribution along the curve that is proportional to the 
Lagrangian Equation of Motion. The extra term also vanishes along 
equal-time curves \mb{x_{0}(t)+t={\rm constant}}.

\setcounter{equation}{0}
\section{Gauge Symmetry}

\vs
\noi
We will now discuss the Hamiltonian formulation.
As mention earlier, our starting point is the Lagrangian
\mb{L[Q](t)} {\em without} assuming chirality of \mb{Q(x,t)}.
The newly gained freedom of the \mb{Q(x,t)}-fields introduces a
gauge symmetry for the the Lagrangian \mb{L[Q](t)} in the following
way; for a given time \mb{t}, let (for the time being) 
\mb{\Sigma_{t}} denote the support
\beq
{\rm supp} \left( x \mapsto
{{\delta L [Q] (t)} \over {\delta Q^{(0)}(x,t)}} \right)
~\equiv~\overline{\left\{ x \in \R \left|~
 {{\delta L [Q] (t)} \over {\delta Q^{(0)}(x,t)}} \neq 0 \right. \right\}}
\label{supp}
\eeq
of the Euler-Lagrange function 
\beq
  x ~~\mapsto~~
{{\delta L [Q] (t)} \over {\delta Q^{(0)}(x,t)}} ~~.
\eeq
It follows from previously made assumptions that \mb{\Sigma_{t}} is
compact. The Lagrangian \mb{L[Q](t)} will be invariant under all 
transformations \mb{\delta Q(x,t)} that leaves \mb{\Sigma_{t}} invariant:
\beq
   \forall x \in \Sigma_{t}~:~ \delta Q(x,t)~=~0~.
\label{gaugesym}
\eeq
The value of the field \mb{Q(x,t)} for \mb{x \notin \Sigma_{t}} has no
physical content. It represents a gauge degree of freedom for the system.

\setcounter{equation}{0}
\section{Boundary Poisson Bracket}

\label{secbpb}

\vs
\noi
We take the boundary Poisson bracket \cite{soloviev,bering,solocomp,family}
for local functionals \mb{F(t)} and \mb{G(t)} to be given by the following 
ultra-local ansatz \cite{family}: 
\beq
\{ F(t), G(t) \} ~=~\sum_{k,\ell=0}^{\infty} c_{k,\ell}
\int_{-\infty}^{\infty} \! \! \! \!d x 
~(\pa_{x})^{k+\ell} \left[{{\delta F (t)} \over {\delta Q^{(k)}(x,t)}}~
{{\delta G (t)} \over {\delta
P^{(\ell)}(x,t)}}\right]~-~(F \leftrightarrow G)~.
\label{bpb}
\eeq
Here \mb{\delta / \delta Q^{(k)}(x,t)} are the higher Euler-Lagrange
derivatives, cf.\ \cite{bering}, and the coefficients \mb{c_{k,\ell}} are 
constants. They are normalized such that \mb{c_{0,0}=1} and such that 
the Jacobi identity is satisfied. In particular, one may show that 
\beq
 c_{k,0}~=~1~=~c_{0,\ell}~.
\label{cknowledge}
\eeq
We can extract the usual canonical equal-\mb{t} relation from \eq{bpb}:
\beq
 \{~ Q(x,t),~ P(x',t)~ \} ~=~ \delta_{\R}(x-x')~.
\label{ccr}
\eeq

\setcounter{equation}{0}
\section{First Class Constraints}

\vs
\noi
The gauge symmetry is generated by the following first class contraints
\beq
 \forall x \notin \Sigma_{t}~:~P(x,t)~ \approx~ 0~~,
\label{hpbc}
\eeq
which is the Hamiltonian version of the boundary condition 
\eq{lpbc}. ( The wavy double line \mb{\approx} is a notation first
introduced by Dirac to denote equality modulo first class constraints,
so-called {\em weak} equality.)
To check in detail that the first class constraint \eq{hpbc} generates
the gauge transformations \eq{gaugesym}, consider the smeared first 
class constraint
\beq
T[\xi](t)~\equiv~ \int_{-\infty}^{\infty} \! \! \! \!d x~\xi(x,t)~ P(x,t)~~,
\eeq
where \mb{\xi(x,t)} is a test function that vanishes on \mb{\Sigma_{t}} :
\beq
  \forall x \in \Sigma_{t}~:~\xi(x,t) ~=~ 0~~.
\eeq
Here \mb{\xi(x,t)} does not depend on the dynamical variables 
\mb{P(x,t)} and \mb{Q(x,t)}.
Let \mb{\delta Q=\xi} be a infinitesimal gauge transformation. 
The gauge variation of the local functional \mb{F[Q,P](t)} is given by 
\bea
\delta_{\xi}F[Q,P](t)&\equiv&F[Q\!+\!\xi,P](t)-F[Q,P](t) \cr\cr
&=&\sum_{k=0}^{\infty}\int_{-\infty}^{\infty} \! \! \! \!d x 
~(\pa_{x})^{k} \left[{{\delta F[Q,P] (t)} \over {\delta Q^{(k)}(x,t)}}~
\xi(x,t)\right]\cr\cr
&=& \{~ F[Q,P](t),~ T[\xi](t)~ \}~. 
\eea
In the last equality, we used \eq{cknowledge}.

\setcounter{equation}{0}
\section{Hamiltonian}

\label{secham}

\vs
\noi
The bare action \mb{S} and the bare Hamiltonian \mb{H(t)} is given by
\bea
S&=& \int_{-\infty}^{\infty} \! \! \! \!dt \left[ 
\int_{-\infty}^{\infty} \! \! \! \!dx ~P(x,t)~ \dot{Q}(x,t) 
~-~H(t)\right] \cr\cr
H(t)&=& \int_{-\infty}^{\infty} \! \! \! \!dx ~P(x,t)~ \pa_{x} Q(x,t) 
~-~L[Q](t)~.
\label{ham}
\eea
The Hamiltonian Equation of Motion 
\beq
 \dot{F}(t)~\approx ~\{ ~F(t),~H(t)~ \}
\eeq
for a local functional \mb{F(t)} becomes
\bea
\lefteqn{
\sum_{k=0}^{\infty}\int_{-\infty}^{\infty} \! \! \! \!d x 
~(\pa_{x})^{k} \left[ {{\delta F (t)} \over {\delta Q^{(k)}(x,t)}}~
\dot{Q}(x,t)+{{\delta F (t)} \over {\delta P^{(k)}(x,t)}}~
\dot{P}(x,t)\right]} \cr\cr
&\approx &\sum_{k=0}^{\infty}\int_{-\infty}^{\infty} \! \! \! \!d x 
~(\pa_{x})^{k} \left[{{\delta F (t)} \over {\delta Q^{(k)}(x,t)}}~
\pa_{x} Q(x,t)
+{{\delta F (t)} \over {\delta P^{(k)}(x,t)}}~\left( \pa_{x} P(x,t)  
+  {{\delta L[Q] (t)} \over {\delta Q^{(0)}(x,t)}}  \right) \right] \cr\cr
&&-\sum_{k=0}^{\infty}c_{1,k}\int_{-\infty}^{\infty} \! \! \! \!d x 
~(\pa_{x})^{k+1} \left[{{\delta F (t)} \over {\delta P^{(k)}(x,t)}}~
 P(x,t)\right]~,
\label{feom}
\eea
where we have used \eq{cknowledge} and the fact that \mb{{\cal L}} 
has compact support. If we furthermore apply the first class constraints
\eq{hpbc}, we can get rid of the last term on the right hand side. 
The remaining equation is clearly equivalent to the chirality 
condition \Eq{qxt} for \mb{Q(x,t)} and \Eq{pdiff} for \mb{P(x,t)}.

\vs
\noi
One may indeed check that the first class constraint
\eq{hpbc} is preserved under the Hamiltonian flow, as it should be:
\bea
 \{ ~T[\xi](t),~H(t)~ \} &=& \int_{-\infty}^{\infty} \! \! \! \!d x
~\xi(x,t) \left( \pa_{x} P(x,t)  
+ {{\delta L[Q] (t)} \over {\delta Q^{(0)}(x,t)}}  \right) ~\approx ~0~,
\cr\cr 
\delta_{\xi}S
 &=& \int_{-\infty}^{\infty} \! \! \! \!d t \left[
- \int_{-\infty}^{\infty} \! \! \! \!d x~\xi(x,t) \dot{P}(x,t)
 +   \{ ~T[\xi](t),~H(t)~ \}  \right] ~\approx ~0~.
\eea
{}Finally, an important remark: We could generalize the previous
discussion by letting \mb{\Sigma_{t}} be {\em any} compact set which 
contains the support \eq{supp}, thereby deliberately choosing a
smaller gauge symmetry. In fact, a transformation \mb{\delta Q(x,t)}
which leaves the support \eq{supp} -- but not the larger set
\mb{\Sigma_{t}} -- invariant, would no longer be a symmetry for the 
action \mb{S} nor the Hamiltonian \mb{H(t)}. The first class
constraint \eq{hpbc} is creating its own justification! This is 
because the original Lagrangian theory -- build out of rigid
left-movers -- does not possess any gauge symmetry at all. 
We see that the theory changes with the choice of \mb{\Sigma_{t}}.
The smaller we choose \mb{\Sigma_{t}}, the more the system is 
prohibited in sending left-moving momenta between different connected
components of the support \eq{supp}. On the other hand, we do not want 
to completely eliminate the gauge symmetry by choosing \mb{\Sigma_{t}}
as large as possible, \ie \mb{\Sigma_{t}=\R}. 
That would complicate matters by activating the extra 
boundary terms appearing in the equations of motion \eq{feom}.
In conclusion, to ensure that our Hamiltonian system corresponds 
to the original Lagrangian theory, we let \mb{\Sigma_{t}} {\em be a compact 
set bigger than the convex hull} of the support \eq{supp}.

\setcounter{equation}{0}
\section{Ostrogradski's Framework}
\label{secostro}

\vs
\noi
Let us recall how the non-local formulation translates 
into Ostrogradski's formulation \cite{ostro} of infinite order.
{}For other treatments, see \cite[Appendix A]{marnelius},
\cite[Section VI A]{llosaold} and \cite[Section 5]{woodardnew}. 
We assume for simplicity that the discontinuity curve
\mb{x_{0}(t)=x_{0}} is constant.
Ostrogradski's coordinates \mb{Q^{(n)}(t)}, \mb{n\in \N_{0}}, 
are defined as
\beq
 Q^{(n)}(t)~=~\left. (\pa_{x})^n Q(x,t) \right|_{x=x_{0}}~.
\eeq 
The inverse relation is given by the Taylor expansion around \mb{x=x_{0}}:
\beq
 Q(x,t)~=~\sum_{n=0}^{\infty}{{(x\!-\!x_{0})^n} \over {n!}}Q^{(n)}(t)~.
\eeq
The chirality condition \Eq{qxt} translates into
\beq
\dot{Q}^{(n)}(t)~=~ Q^{(n+1)}(t) ~.
\label{oqxt}
\eeq

\vs
\noi
To ensure that the boundary Poisson bracket \Eq{bpb} 
corresponds to the discrete analogue given by
\beq
\{ Q^{(n)}(t),P_{(m)}(t) \} ~=~ \delta^{n}_{m}~,
\eeq
we define Ostrogradski's momenta \mb{P_{(n)}(t)} as
\beq
 P_{(n)}(t)~=~\int_{-\infty}^{\infty} \! \! \! \! dx ~ 
{{(x\!-\!x_{0})^n} \over {n!}}~ P(x,t)~.
\eeq
The integral is well-defined because the momenta \mb{P(x,t)} have 
compact support, cf.\ \Eq{lpbc}. The inverse relation reads  
\beq
 P(x,t)~=~\sum_{n=0}^{\infty}P_{(n)}(t) ~(-\pa_{x})^n  
\delta_{\R}(x\!-\!x_{0}) ~.
\eeq
Alternatively, the formulas for the momenta follows from the 
Schr\"{o}dinger representation
\beq
 {{\delta} \over {\delta Q^{(0)}(x,t)}}~=~\sum_{n=0}^{\infty}
 (-\pa_{x})^n  \delta_{\R}(x\!-\!x_{0})~ {{\pa} \over {\pa Q^{(n)}(t)}}~,
\eeq
and equivalently
\beq
{{\pa} \over {\pa Q^{(n)}(t)}} ~=~
\int_{-\infty}^{\infty} \! \! \! \! dx ~ {{(x\!-\!x_{0})^n} \over {n!}}~ 
{{\delta} \over {\delta Q^{(0)}(x,t)}}~.
\eeq
The equations \eq{pdef} and \eq{pdiff} translate into 
\beq
 P_{(n)}(t)~=~\sum_{m=n}^{\infty} (-\pa_{t})^{m-n} 
{{\pa L[Q](t)} \over {\pa Q^{(m+1)}(t)}}
\eeq
and
\beq
\left\{ \begin{array}{rcl}
 \dot{P}_{(n)}(t)+ P_{(n-1)}(t)
&=&{{\pa L[Q](t)} \over {\pa Q^{(n)}(t)}}~,~~n \in \N~, \cr\cr
\dot{P}_{(0)}(t)&=&{{\pa L[Q](t)} \over {\pa Q^{(0)}(t)}}~,
\end{array}
\right.
\label{opdiff}
\eeq
respectively. The very last equation is the Lagrangian Equation of Motion.
The Hamiltonian \eq{ham} translates into 
\beq
H(t)~=~\sum_{n=0}^{\infty} P_{(n)}(t)~Q^{(n+1)}(t) 
~-~ L[Q](t)~.
\label{oham}
\eeq
The Hamiltonian Equations of Motion are \eq{oqxt} and \eq{opdiff}.

\setcounter{equation}{0}
\section{Discrete Case}
\label{dica}

\vs
\noi 
{}Finally, let us consider the discrete case, cf.\ Section~\ref{lft}, 
with constant curves \mb{x_{i}(t)=x_{i}}.  
The Euler-Lagrange equation for \mb{Q} reads
\bea
{{\delta L [Q] (t)} \over {\delta Q^{(0)}(x,t)}}
&=&\sum_{i \in I} \sum_{k=0}^{\infty} 
{{\pa L [Q] (t)} \over {\pa Q^{(k)}(x_{i},t)}}~ 
(-\pa_{x} )^{k} \delta_{\R}(x\!-\!x_{i}) \cr\cr
&=&\sum_{i \in I} \sum_{k=0}^{\infty} 
E_{(k)}(x_{i},t\!+\!x\!-\!x_{i})~
(-\pa_{x} )^{k} \delta_{\R}(x\!-\!x_{i})~.
\eea
Here we have introduced the higher Euler-Lagrange derivatives, 
cf.\ \cite{bering}:
\beq
 E_{(k)}(x,t)~=~ \sum_{m=k}^{\infty}\twobyone{m}{k}
(-\pa_{t} )^{m-k} 
{{\pa L [Q] (t)} \over {\pa Q^{(m)}(x,t)}}
~~,~~~~~~x=x_{i}~~,~~~~~k \in \Z~~.
\eeq
An alternative basis is provided by the Ostrogradski derivatives:
\beq
 O_{(k)}(x,t)~=~ \sum_{m=k}^{\infty}
(-\pa_{t} )^{m-k} 
{{\pa L [Q] (t)} \over {\pa Q^{(m)}(x,t)}}
~~,~~~~~~x=x_{i}~~,~~~~~k \in \Z~~.
\eeq
The partial derivatives can be recovered via the inverse relation
\beq
{{\pa L [Q] (t)} \over {\pa Q^{(k)}(x,t)}}
~=~O_{(k)}(x,t)+\pa_{t} O_{(k+1)}(x,t)
~~,~~~~~~x=x_{i}~~,~~~~~k \in \Z~~.
\eeq
We assume that these are regular \mb{C^{\infty}}-functions. Note 
the peculiar fact, that although the Lagrangian \mb{L[Q](t)} contains no
temporal derivatives \mb{\pa_{t}} -- only spatial derivatives -- it is the
temporal derivatives \mb{\pa_{t}} that is used in the construction of
the above functions. Of course, this is the same on-shell, \ie when
\eq{qxt} holds.

\vs
\noi
The momentum formula \eq{pdef} becomes:
\bea
P(x,t)&=& \theta(x\!-\!x_{0}(t)) 
\sum_{k=0}^{\infty}  (-\pa_{x})^{k}\sum_{i \in I}
\left[\theta(x_{i}\!-\!x)~ 
{{\pa L [Q] (t\!+\!x\!-\!x_{i})} \over
 {\pa Q^{(k)}(x_{i},t\!+\!x\!-\!x_{i})}} \right] \cr\cr 
&&-~\theta(x_{0}(t)\!-\!x) 
\sum_{k=0}^{\infty} (-\pa_{x} )^{k}\sum_{i \in I}
\left[\theta(x\!-\!x_{i})~ 
{{\pa L [Q] (t\!+\!x\!-\!x_{i})} \over
 {\pa Q^{(k)}(x_{i},t\!+\!x\!-\!x_{i})}} \right]\cr\cr 
&=& \sum_{i \in I}\sum_{k=0}^{\infty}  
E_{(k)}(x_{i},t\!+\!x\!-\!x_{i})~ 
\left[  \theta(x\!-\!x_{0}(t)) ~(-\pa_{x} )^{k}
 \theta(x_{i}\!-\!x) -   \theta(x_{0}(t)\!-\!x)~
(-\pa_{x} )^{k}  \theta(x\!-\!x_{i})  \right] \cr\cr 
&=& \sum_{i \in I} \sum_{k=1}^{\infty}
E_{(k)}(x_{i},t\!+\!x\!-\!x_{i})~
(-\pa_{x})^{k-1}\delta_{\R}(x\!-\!x_{i})  \cr\cr 
&&+\sum_{i \in I} E_{(0)}(x_{i},t\!+\!x\!-\!x_{i})~ 
\left[ \theta(x_{i}\!-\!x)~ \theta(x\!-\!x_{0}(t))
-  \theta(x_{0}(t)\!-\!x)~ \theta(x\!-\!x_{i}) \right]    \cr\cr 
&=& \sum_{i \in I} \sum_{k=1}^{\infty} O_{(k)}(x_{i},t)~
 (-\pa_{x} )^{k-1}\delta_{\R}(x\!-\!x_{i}) \cr\cr
&&+\sum_{i \in I} O_{(0)}(x_{i},t\!+\!x\!-\!x_{i})~  
\left[ \theta(x_{i}\!-\!x)~ \theta(x\!-\!x_{0}(t))
-  \theta(x_{0}(t)\!-\!x)~ \theta(x\!-\!x_{i}) \right] 
   ~. 
\label{pdefdiscr}
\eea
We learn that the typical momenta for a discrete system will be 
distributional in nature with non-smooth behavior at the discrete points
\mb{x=x_{i}}. Note that the support of the momenta
\mb{x \mapsto P(x,t)} is inside the convex hull of the support
\eq{supp} (= \mb{\{ x_{i} | i \in I \}}), if \mb{x_{0}(t)} is.

\vs
\noi
The Lagrangian Equation of Motion for \mb{q}, cf.\ \Eq{leom}, reads:
\beq
0~=~{{\delta S[q]} \over {\delta q^{(0)}(x\!+\!t)}}
~\equi{\eq{qxt}}~  \sum_{i \in I}  E_{(0)}(x_{i},t\!+\!x\!-\!x_{i}) ~.
\label{dleom} 
\eeq

\setcounter{equation}{0}
\section{Discrete Local Formulation}

\vs
\noi
We continue to study the discrete case.
We have already given a local field formulation above. But after all, 
it is awkward to use field theory to quantize a discrete system!
It is natural to ask if one can get a local formulation {\em without} 
introducing fields? Ostrogradski's formal approach, 
cf.\ Section~\ref{secostro}, is cumbersome in practice 
if the index set \mb{I} contains more than one element. 

\vs
\noi If we wish to address the original problem -- and hence 
choosing \mb{\Sigma_{t}} to be the convex hull of 
\mb{\{ x_{i} | i \in I \}} -- there seems to be 
no easy/useful/natural local discrete formulation in general. 

\vs
\noi
However, if we let \mb{\Sigma_{t}} be the smallest set possible, 
\ie \mb{\{ x_{i} | i \in I \}}, we can do better. 
The first class constraints \eq{hpbc} along with the equations of motion 
\eq{pdiff} would then exclude the last term involving the Heaviside 
step function \mb{\theta} in the last two expressions of \eq{pdefdiscr}.
It implies an Euler-Lagrange equation for each individual \mb{i \in I}: 
\beq
    O_{(0)}(x_{i},t)~\equiv~  E_{(0)}(x_{i},t)~=~0~.
\eeq
This corresponds to a theory {\em without} the chirality condition \eq{qxt}.
We can fit this inside a generalized Ostrogradski framework with new 
coordinates \mb{Q^{i(n)}(t)}, \mb{i \in I},  \mb{n\in \N_{0}}, defined as
\beq
 Q^{i(n)}(t)~=~\left. (\pa_{x})^n Q(x,t) \right|_{x=x_{i}}~.
\eeq 
and momenta \mb{P_{i(n)}(t)}, \mb{i \in I}, \mb{n\in \N_{0}}, 
 \beq
P_{i(n)}(t)~=~\lim_{\epsilon \to 0^{+}} 
\int_{x_{i}-\epsilon}^{x_{i}+\epsilon} \! \! \! \! dx ~ 
{{(x\!-\!x_{i})^n} \over {n!}}~ P(x,t)~,
\eeq
canonical Poisson bracket
\beq
\{ Q^{i(n)}(t),P_{j(m)}(t) \} ~=~ \delta^{n}_{m}~\delta^{i}_{j}~,
\eeq
and Hamiltonian
\beq
H(t)~=~\sum_{i \in I} \sum_{n=0}^{\infty} P_{i(n)}(t)~Q^{i(n+1)}(t) 
~-~ L[Q](t)~.
\eeq            
More relations can be read off from the Section~\ref{secostro} with the 
obvious minor modifications. In this way we have obtained an unconstrained 
discrete Hamiltonian system with slightly generalized Ostrogradski 
coordinates. 
For instance, the classical momentum formula \eq{pdefdiscr} turns into
\beq
   P_{i(n)}(t)~=~O_{(n+1)}(x_{i},t)
~~, ~~~~~i \in I~~, ~~~~~~n\in \N_{0}~~. 
\eeq


\vs
\noi
{\bf Acknowledgements}.
We would like to thank R.P.~Woodard and B.D.~Baker for comments.
The research is supported by DoE grant no.~DE-FG02-97ER-41029.



\end{document}